\documentclass[%
 notitlepage,
 reprint,
 superscriptaddress,
 amsmath,amssymb,
 aps,
]{revtex4-1}

\usepackage{lipsum}
\usepackage{graphicx}% Include figure files
\usepackage{dcolumn}% Align table columns on decimal point
\usepackage{bm}% bold math
\usepackage{epstopdf}

\newcommand{\bra}[1]{\langle #1 \vert}
\newcommand{\ket}[1]{\vert #1 \rangle}

\begin{document}

\preprint{APS/123-QED}

\title{Disorder-assisted distribution of entanglement in $XY$ spin chains}

\author{Guilherme M. A. Almeida}
\email{gmaalmeida.phys@gmail.com}
\affiliation{%
 Instituto de F\'{i}sica, Universidade Federal de Alagoas, 57072-900 Macei\'{o}, AL, Brazil
}%
\author{Francisco A. B. F. de Moura}
\affiliation{%
 Instituto de F\'{i}sica, Universidade Federal de Alagoas, 57072-900 Macei\'{o}, AL, Brazil
}%
\author{Tony J. G. Apollaro}%  

\affiliation{Quantum Technology Lab, Dipartimento di Fisica, Universit$\grave{a}$ degli Studi di Milano, 20133 Milano, Italy}
\affiliation{Centre for Theoretical Atomic, Molecular, and Optical Physics, School of Mathematics an Physics, Queen's University Belfast, BT7,1NN, United Kingdom}
\author{Marcelo L. Lyra}
\affiliation{%
 Instituto de F\'{i}sica, Universidade Federal de Alagoas, 57072-900 Macei\'{o}, AL, Brazil
}%

\date{\today}% It is always \today, today,

\begin{abstract}
We study the creation and distribution of entanglement in 
disordered $XY$-type spin-$1/2$ chains
for the paradigmatic case of a single flipped spin prepared on a 
fully polarized background.
The local magnetic field is set to follow a disordered long-range-correlated sequence with power-law spectrum. Depending on the degree of correlations of the disorder, a set of extended
modes emerge in the middle of the band yielding an
interplay between localization and delocalization.
As a consequence,
%which brings about
a rich variety of entanglement distribution patterns arises, which we evaluate here through the
concurrence between two spins. 
%featuring a single tunable parameter. 
%This allows for moving from the uncorrelated disorder scenario, 
%at which the spin is unable to propagate
%due to the emergence of strongly localized states,
%to a regime where a portion of the spectrum is populated by extended states.
%
We show that, even in the presence of disorder,
the entanglement wave can be pushed to spread out
reaching distant sites and also enhance 
pairwise entanglement between the initial site 
and the rest of the chain. 
We also study the propagation of 
an initial maximally-entangled state 
through the chain and show that correlated disorder improves the transmission
quite significantly when compared with the uncorrelated counterpart. 
%Moreover, the degree of correlation ultimately controls the 
%distribution profile. 
%
%Thereby, imposing internal correlations in the disordered 
%sequence becomes an alternative to overcome defects that may arise in the fabrication
%of spin chains. 
%thus offering further alternatives for 
%quantum-information-processing control in more realistic settings.
Our work contributes in designing solid-state devices for quantum information processing in the realistic setting of correlated static disorder.  

%\begin{description}
% Enter here
%\end{description}
\end{abstract}

\maketitle

%\tableofcontents

\section{\label{sec1}Introduction}

% SPIN CHAINS
In the last few decades,  
since the seminal proposal put forward by Bose \cite{bose03},
much attention has been given to solid-state hardware where information is 
encoded in stationary spins acting as {\it qubits} in which the energy splitting is induced by a local magnetic field 
and the (usually nearest-neighbor) coupling between them is set by their exchange interaction. 
Following that, it has been shown that low-dimensional spin chains can act as efficient (especially for short-distance communication)
quantum ``wires'' 
%for transmitting quantum states [...] and entanglement [...].  
for carrying out quantum-state transfer protocols 
\cite{bose03, christandl04,plenio04,wojcik05,li05,huo08, gualdi08, banchi10, *banchi11,apollaro12,lorenzo13,lorenzo15} 
as well as creation 
%e.g. cluster states
and distribution of entanglement \cite{amico04, apollaro06,plastina07,*apollaro08, hartmann06ent,tsomokos07,venuti07, cubitt08, giampaolo09, *giampaolo10, gualdi11, estarellas17, *estarellas17scirep}, 
both being pivotal tasks in quantum networks \cite{cirac97}.
%
% EXPERIMENTAL IMPLEMENTATIONS
%Spin chains can be (and have been) readly implemented in
Physically, spin chains may be implemented in
many platforms such as  
NMR systems \cite{cappellaro07, *alvarez10prl},
optical lattices \cite{weitenberg11, fukuhara13, *fukuhara15},
arrays of coupled cavity-QED systems \cite{angelakis07, almeida16},
superconducting circuits \cite{romito05, *johnson11},  
nitrogen vacancies in diamond \cite{ping13},
and waveguides \cite{bellec12}.

% ADVANTAGES 
The main advantages of using spin chains as quantum channels are twofold. First, 
they bypass
the need for inter-converting between photons and qubits, 
such as in hybrid light-matter devices \cite{cirac97, almeida13, almeida16}, 
which demands a high degree of control and 
may lead to decoherence and losses.
%There is also the fact that photons weakly interact with other stuff and among themselves. 
Moreover, most of the protocols require minimal user control (mostly at the sender and receiver sites) 
as the system's dynamics is driven through 
the evolution of the underlying Hamiltonian, offering thus a versatile toolbox for quantum information processing purposes.
%
%That is where the versatility offered by spin chains lies in. 
%Their Hamiltonian
%can be simple enough so that one can come up with a  
%predefined task and engineer the system couplings accordingly.
The analytic tractability of the spin Hamiltonian has allowed for several theoretical investigations.
For instance, a specific modulation of the entire chain allows for perfect state transfer
as shown in Refs. \cite{christandl04,plenio04} (see also \cite{feder06}). 
By adding in local defects, either in form of magnetic fields or coupling strengths, it is possible 
to carry out high-fidelity quantum-state transfer (it takes place by effectively reducing the operating Hilbert space) \cite{wojcik05, huo08, kuznetsova08, gualdi08, lorenzo13,lorenzo15}, routing protocols \cite{wojcik07, ross11, paganelli13}, 
and to control and enhance entanglement distribution \cite{apollaro06,plastina07, *apollaro08}. 
Also, it has been shown that
gapped dimerized models plays a major role in 
establishing 
long-distance communication \cite{venuti07,huo08, kuznetsova08,giampaolo09, *giampaolo10, almeida16, estarellas17, *estarellas17scirep}.
%
%Thus, inhomogeneities in general, offer 
%a great deal of opportunities for quantum information processing. 
 
% IMPORTANT PART
All these 
essential tasks
that can be performed using spin chains
rely on how precisely its parameters
can be tuned. 
%e.g., exchange couplings and magnetic field.
Thereby, the very task of engineering those systems poses out
a few challenging issues.
First of all, it is not trivial to address a single spin with the desired precision 
although significant progress has been achieved \cite{weitenberg11}.
%Also, the environment may cause decoherence and other kinds of noise. 
%
Another crucial one is due to fabrication errors (e.g. spin positioning) which 
lead to disorder and thus localization of quantum information \cite{burrell07, *allcock09, lorenzo17}.
% 
%Nevertheless, we may ask ourselves: how crucial is disorder?
Since experimental imperfections may always be present
in solid-state devices, it becomes essential 
to analyze their robustness to that and check whether it is even possible to enhance
transfer of information through noisy channels 
\cite{allcock09, dechiara05,fitzsimons05,burgarth05,tsomokos07, petrosyan10,yao11,zwick11,*zwick12, bruderer12,kay16}.    
Most of the works in this direction have addressed the effects of
site-independent random fluctuations (static disorder) 
and the general verdict is that some protocols are prone
to survive to it as long as disorder does not overcome a given threshold.

It is well established that single-particle eigenstates 
of 1D tight-binding models featuring random potentials are all exponentially localized no matter
how strong disorder is \cite{anderson58,*abrahams79}. 
This is, however, no longer true when internal correlations in the disorder distribution are present.  
It was shown that short-range spatial correlations
induce the breakdown of Anderson localization, thereby elucidating 
transport properties for a wide class of polymers \cite{dunlap90, *phillips91}.
Following the road, it was demonstrated that long-range correlations \cite{demoura98,izrailev99} 
promote the appearance of a band of extended states with well defined
mobility edges separating them from localized states thus revealing 
an Anderson-like metal-insulator transition.
That was later confirmed using single-mode waveguides \cite{kuhl00} and
many related experiments have been carried out since then \cite{kuhl08, *dietz11}   
(see \cite{izrailev12rev} for a recent review on the subject).
Right after the above findings took place, 
there has been a tremendous interest in 
investigating dynamical properties of various 1D models featuring 
either diagonal (on-site potential)
or off-diagonal (hopping strength) correlated disorder \cite{lima02, *demoura02,*nunes16,demoura03,adame03, laguna16}. 
Particularly, it was shown that the one-magnon spectrum of 
ferromagnetic chains \cite{lima02} 
can exhibit a phase of 
low-energy delocalized states thus     
displaying a rich set of dynamical regimes. 

Generally speaking, 
the interplay between localized and delocalized states sets
a suitable ground for designing quantum information processing protocols and that is exactly
where our work fits in. Here we 
address how long-range correlated diagonal disorder affects the transport of
quantum-information in spin chains. Specifically, 
we aim to track down
the entanglement wave produced after a tiny disturbance on the fully-polarized state, that is a single flipped spin
set in the bulk of an isotropic $XY$ chain. 
This is a rather paradigmatic scenario and has been tackled  
in recent experiments \cite{fukuhara13, *fukuhara15}. 
%We formally evaluate entanglement
%via the entropy for a block of spins, analyzing how it scales with the size of the block \cite{vidal03}, and we
%also verify the amount of pairwise entanglement along the chain quantified by the concurrence \cite{wootters98}. 
%

In this work,
the local magnetic field is set to follow a disorder distribution
with power-law spectrum of the form $S(k) \propto 1/k^{\alpha}$, where $k$ is the corresponding wave number, and
$\alpha$ accounts for the degree of long-range correlations. 
We emphasize that such kind of long-range correlated disorder must not be viewed solely as an imposed scenario on the spin chain.
In fact, 
many stochastic processes 
in nature featuring long-range correlations are expected to obey
a power-law spectrum such as, to name a few examples, nucleotide sequences in DNA molecules \cite{peng92, *carpena02},
which may strongly affect electronic transport, plasma fluctuations \cite{carreras98}, and patterns in surface growth \cite{lam92}.
Therefore, on the one hand, fluctuations arising from
the fabrication of
solid-state quantum information devices might not always be completely random, i.e., uncorrelated.
On the other one, instead of trying to impose a set of finely-tuned parameters,
it should be more realistic, on the experimental side, to devise a scheme that
allows for the presence of correlated disorder. 
%
%Surprisingly, this very detail can be achieved with disorder itself, as long as 
%its underlying correlations can be tailored. 
%

Here, we show that 
the \textit{degree} of 
long-range correlated disorder, $\alpha$, ultimately controls  
the entanglement distribution profile after the excitation
is released from the middle of the chain following the system's Hamiltonian dynamics.
Moreover, we report an
enhancement of entanglement between 
the initial site and the rest of the chain 
as $\alpha$ goes from zero (uncorrelated disorder) to $\alpha=2$. 
For $\alpha>2$, entanglement becomes more prominent 
between symmetrically located spins,
with respect to the initial site, and their nearest neighbors. 
Furthermore, we study the propagation of a Bell-type state 
through the chain and show that the entanglement transmission 
coefficient (through a given fixed site of the chain) 
significantly increases with $\alpha$ to the point of
surpassing the reflection coefficient.

%Furthermore, by defining entanglement transmission and reflection coefficients
%based on the saturation of the  
%Coffman-Kundu-Wootters conjecture \cite{coffman00} for
%single-particle states, we study %the propagation of 
%an initial maximally-entangled state across the chain and show 
%the transmission can be improved

The remainder of this paper is organized as follows. In Sec. \ref{sec2}, we 
introduce the spin Hamiltonian featuring on-site long-range correlated disorder. 
In Sec. \ref{sec3}, we briefly discuss the entanglement measurements which will be used through the paper. In Sec. \ref{sec4} we discuss the dynamics of 
entanglement for the ordered chain and show our results 
for the disordered scenario. 
Final comments are addressed in Sec. \ref{sec5}.

\section{\label{sec2}Spin-chain Hamiltonian}

% FIG 1
\begin{figure}[t!] 
\includegraphics[width=0.42\textwidth]{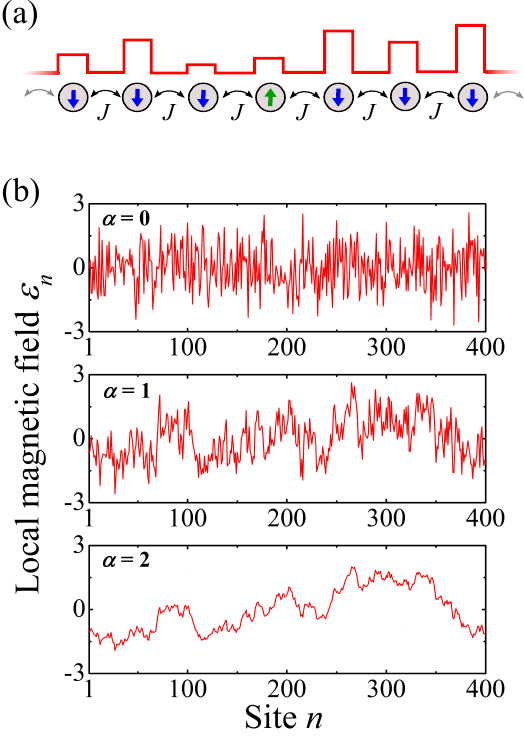}
\caption{\label{fig1} 
(Color online)
(a) Sketch of the spin chain featuring XX-type exchange interactions
with strength $J$ subjected to a random on-site potential landscape (depicted by red bars).
The spin chain is initialized in a fully-polarized state and a single excitation 
(a flipped spin) is set, say, in the middle of it. We thus let it to evolve via its Hamiltonian dynamics. 
(b) Single realization of the disorder distribution $\lbrace \varepsilon_{n} \rbrace$ (in units of $J$) generated from Eq. (\ref{disorder}) for $\alpha = 0$, $1$, and $2$, and $N = 400$.
The entire sequence is normalized satisfying $\langle \epsilon_{n} \rangle = 0$ and $\langle \epsilon_{n}^2 \rangle = 1$.
We note that by increasing $\alpha$ 
the distribution smooths out resembling the trace of a fractional Brownian motion (see the last panel for $\alpha=2$).  
}
\end{figure}

We consider an one-dimensional isotropic spin chain featuring XX-type exchange interactions as given by the Hamiltonian ($\hbar = 1$)
\begin{equation} \label{spinH}
\hat{H}_{S} = \sum_{n=1}^{N}\dfrac{\varepsilon_{n}}{2}(\hat{1}-\hat{\sigma}_{n}^{z})-\sum_{n=1}^{N-1}\dfrac{J_{n}}{2}(\hat{\sigma}_{n}^{x}\hat{\sigma}_{n+1}^{x}
+\hat{\sigma}_{n}^{y}\hat{\sigma}_{n+1}^{y}), 
\end{equation}
where $\hat{\sigma}_{n}^{x,y,z}$ are the usual Pauli operators for the $n$-th spin, $\varepsilon_{n}$ is the strength of the local magnetic field, 
and $J_{n}$ is the exchange coupling rate. The above Hamiltonian can be put into another equivalent form through the Jordan-Wigner transformation which maps the spin Hamiltonian
(\ref{spinH}) onto a system of non-interacting spinless fermions 
\begin{equation}\label{fermionH}
\hat{H} = \sum_{n=1}^{N}\varepsilon_{n} \hat{c}_{n}^{\dagger}\hat{c}_{n} - 
\sum_{n=1}^{N-1}  
J_{n}\left( \hat{c}_{n}^{\dagger}\hat{c}_{n+1}+\hat{c}_{n+1}^{\dagger}\hat{c}_{n} \right),
\end{equation}
where $\hat{c}_{n}^{\dagger}$ ($\hat{c}_{n}$) creates (annihilates) a particle
at the $n-$th site. 
%Those operators obey the standard anticommutation rules. 
That way, the presence (absence) of a fermion in a given site represents 
a spin up (down) state.
Note that since $\left[ \hat{H}, \sum_{n} \hat{c}_{n}^{\dagger}\hat{c}_{n}\right] =0$,
Hamiltonian (\ref{fermionH}) can be split into number-invariant subspaces. 
Here, we aim to study
the entanglement 
generated from a single particle prepared in a given location
on top of the 
fully-polarized state $\ket{\mathrm{vac}}\equiv \ket{00...0}$ [see Fig \ref{fig1}(a)].  
Thereby, the whole dynamics takes place in the single-excitation sector 
$\left\lbrace \ket{n} \right\rbrace $, 
with $\ket{n} \equiv \hat{c}_{n}^{\dagger} \ket{\mathrm{vac}}$ and Eq. (\ref{fermionH}) takes the form of a $N\times N$ tridiagonal hopping matrix. 

Now, let us make a few considerations about the parameters of the chain. First, 
we set a uniform distribution of hopping rates (exchange interaction) 
$J_{n} = J$.
The on-site potentials (local magnetic fields), however, 
will follow a disorder distribution of a special kind.
Disorder can arise from manufacturing imperfections and/or due to 
dynamical factors. Either way, we can safely assume that the noise is static, that is
it does not change considerably over time.
Here we consider random sequences featuring long-range spatial correlations and
one of the most simple and convenient ways to model it 
is by considering the trace of a fractional Brownian motion with power-law
spectrum $S(k)\propto1/k^{\alpha}$ which can be built from \cite{demoura98, adame03}
%where $S(k)$ is the Fourier transform 
%of the correlation function $\langle \varepsilon_{i}\varepsilon_{j} \rangle$
\begin{equation} \label{disorder}
\varepsilon_{n} = \sum_{k=1}^{N/2}k^{-\alpha/2}
\mathrm{cos}\left( \dfrac{2\pi n k}{N} + \phi_{k} \right),
\end{equation}  
where $k=1/\lambda$ with $\lambda$  being the wavelength 
of the modulation profile, $\lbrace \phi_{k} \rbrace$ are random
phases uniformly distributed within $\left[ 0,2\pi \right]$, 
and $\alpha$ ultimately accounts for the degree of correlations. Hereafter we normalize the 
above sequence to have zero mean and unit variance which can be done by simply
redefining $\varepsilon_{n} \rightarrow 
\left( \varepsilon_{n}-\langle\varepsilon_{n}\rangle \right) / \sqrt{\langle\varepsilon_{n}^2\rangle-\langle\varepsilon_{n}\rangle^2}$. 
%This must be done in order to keep... \textbf{[Francisco??]}. 
We emphasize that the disorder distribution generated by the  above formalism has no typical length scale, which is a characteristic of several stochastically generated natural series\cite{bak96}. The power-law spectral density $S(k)\propto 1/k^{\alpha}$  is a direct  consequence of the power-law form of the two-point correlation function. Indeed, $\alpha$ is related with the Hurst exponent \cite{fractalsbook} through  
$H = (\alpha -1)/2$.  This quantity  describes the self-similarities of the series and also the persistence  of its increments.  
For $\alpha= 0$, the sequence is completely uncorrelated, whereas for any value  $\alpha>0$ intrinsic long-range correlations appears.  The value $\alpha =2$ represents  the case in which the sequence mimics  the trace of a Brownian motion. When $\alpha > 2$ ($\alpha < 2$) the increments on the series are said to be persistent (anti-persistent).

In Fig. \ref{fig1}(b)
we show samples generated by Eq. (\ref{disorder}) for different values of $\alpha$.
For $\alpha=0$ we recover the standard uncorrelated disorder (white noise)
distribution
where $\langle \varepsilon_{i}\varepsilon_{j} \rangle =  \langle \varepsilon_{i}^{2} \rangle \delta_{i,j}$. 
For $\alpha > 0$, internal correlations take place
giving rise to the trace of a Brownian motion when $\alpha = 2$. The sequence
 becomes less rough as we further increase 
the parameter $\alpha$ \cite{demoura98}.
Interestingly, it was shown in \cite{demoura98} 
for a tight-binding electronic model
that the fractal nature of the potential landscape arising from Eq. (\ref{disorder}) 
dictates the appearance of delocalized electronic states 
around the band center 
of the one-particle spectrum 
when $\alpha > 2$.
% This further indicates that a Anderson-like metal to insulator transition
%takes place
%
Those are kept apart from localized states by two mobility edges. 
Similar behavior was also reported in Refs. \cite{izrailev99, demoura02,lima02, demoura03}.

Here we investigate how the interplay between localized and delocalized modes
affects the dynamics of entanglement 
in XY spin chains described by Hamiltonian (\ref{fermionH}) 
which has the form of a standard hopping model.  
% after we place an elementary spin excitation on top of the fully polarized state
%and discuss how it can be used.......... 
Right before that, let us first
illustrate the tools we will adopt to quantify entanglement.

%participation ratio?

\section{\label{sec3}Quantifying entanglement}

Here we deal with two common measures of bipartite entanglement, namely the von Neumann entropy
which addresses the amount of entanglement a given subsystem (say, a spin block)
is sharing with the rest the chain (the whole system being in a pure state) and the so-called 
concurrence \cite{wootters98} which 
is the most suitable tool for characterizing entanglement
between two qubits in an arbitrary mixed state. 

Let us consider a \textit{single} quantum particle hopping on a 
network with $N$ sites modelled by a Hamiltonian
of the form of Eq. (\ref{fermionH}). Note that this hopping particle 
may represent an \textit{actual} fermion or boson, or, which is our case, 
a single flipped spin propagating along the 
chain via exchange interactions [c.f. Eqs. (\ref{spinH}) and (\ref{fermionH})].
Generally speaking, whenever we mention \textit{qubit}, we mean the
two logical states $\ket{0_{i}}$ and $\ket{1_{i}}$ corresponding
to the eigenstates of $\hat{\sigma}_{i}^{z}$. Furthermore, because of the conservation
of the total magnetization in the $z$-direction and the presence of at most one
flipped spin in the chain, $\ket{0_{i}}$ ($\ket{1_{i}}$) matches with the absence (presence) of a fermion at the $i-$th site.
%absence $\ket{0_{j}}$ or presence $\ket{1_{j}}$ of the particle at a given site $j$.
%
%The approach below is valdid for any single-particle set of states defined 
%on an arbitrary network  
  
Any given arbitrary state in the single-particle sector can be written as a linear combination of the single-excitation
basis $\lbrace \ket{i} \rbrace$, that is
\begin{equation}
\ket{\psi} = \sum_{i}w_{i}\ket{i},
\end{equation}
with $w_{i}$ being a complex coefficient in
such a way that $\vert w_{i} \vert^{2}$ is the probability of finding the
particle at site $i$.
In the density operator formalism, it reads
\begin{equation}
\rho = \ket{\psi}\bra{\psi} = \sum_{i}\sum_{j}w_{i}w_{j}^{*}\ket{i}\bra{j}.
\end{equation}

Now suppose we want to write down the state for a block of spins $\mathcal{A}_{L}$ of size $L$. This can be done by choosing a specific set of sites and 
\textit{tracing out} the rest of them, $\rho_{L} = \mathrm{Tr}_{\mathcal{B}_{N-L}}\rho$,
where $\mathcal{B}_{N-L}$ denotes the remaining set. The resulting reduced density operator, expressed in its diagonal basis, is given by $\rho_{L}=\mathrm{diag}[p,1-p]$, where $p \equiv \sum_{i \in \mathcal{A}_{L}}\vert w_{i} \vert^{2}$.
A quite straightforward way to compute the entanglement between
both partitions $\mathcal{A}$ and $\mathcal{B}$, given that the overall state is pure, is 
through the well-known von Neumann
entropy 
%often referred to as entropy of entanglement [Phys. Rev. A 53, 2046 (1996)]
\begin{equation}\label{entropy}
S[\rho_{L}] = -\mathrm{Tr}\rho_{L}\mathrm{log}_{2}\rho_{L}=
-p\mathrm{log}_{2}p-(1-p)\mathrm{log}_{2}(1-p)
\end{equation}
which in our case is bounded by the interval $[0,1]$, with $0$
accounting for a product (separable) state and $1$ for a fully-entangled one.
%Note that we could also have used the other subsystem (this is kinda obvious...)
The entropy above thus depends only on the total probability $p$ of finding the excitation
within block $\mathcal{A}_{L}$, reaching its maximum when $p=1/2$.

In order to characterize how much entanglement can be found in a given
pair of spins, say $i$ and $j$, we once again evaluate the reduced density operator 
which, 
%now yields a 
%four dimensional matrix spanned 
in the basis $\lbrace \ket{0_{i}0_{j}}, \ket{1_{i}0_{j}}, \ket{0_{i}1_{j}}, \ket{1_{i}1_{j}} \rbrace$, reads
\begin{equation} \label{rhoij}
\rho_{i,j}=
\begin{bmatrix}
1-|w_{i}|^{2}-|w_{j}|^{2} & 0 & 0 & 0\\ 
 0& |w_{i}|^{2} & w_{i}w_{j}^{*} & 0\\ 
 0&w_{j}w_{i}^{*}  & |w_{j}|^{2} & 0\\ 
 0& 0 & 0 & 0
\end{bmatrix}. 
\end{equation}
The two-site reduced density matrix above is all we need  
to evaluate the amount of entanglement shared by the pair of spins
through the concurrence \cite{wootters98}
which, given a general bipartite mixed state $\rho_{AB}$ of two qubits, 
is defined by
\begin{equation}
C(\rho_{AB}) = \mathrm{max}\lbrace 0, \sqrt{\lambda_{1}}-\sqrt{\lambda_{2}}-\sqrt{\lambda_{3}}-\sqrt{\lambda_{4}} \rbrace,
\end{equation}
where $\lbrace \lambda_{i} \rbrace$ are the eigenvalues, in decreasing order of the non-Hermitian matrix $\rho_{AB} \tilde{\rho}_{AB}$, where
\begin{equation}
\tilde{\rho}_{AB} = (\hat{\sigma}_{y}\otimes \hat{\sigma}_{y})\rho_{AB}^{*}(\hat{\sigma}_{y}\otimes \hat{\sigma}_{y})
\end{equation}
%being the ``spin-flipped'' matrix, 
and $\rho_{AB}^{*}$ is the complex conjugate of $\rho_{AB}$. 
%and $\sigma_{y}$ the Pauli operator.
For separable qubits, we have $C = 0$. On the other hand, for fully-entangled particles, $C=1$.
In our case, Eq. (\ref{rhoij}), the concurrence reads
\begin{equation}\label{c_ij}
C_{i,j} \equiv C(\rho_{i,j}) = 2\vert w_{i}w_{j}\vert,
\end{equation}
that is, it depends only on the wave-function amplitude of both
spins of interest. 

\section{\label{sec4}Time evolution of entanglement}

In this section, we investigate the time evolution of entanglement 
in the spin chain described by Hamiltonian (\ref{spinH}) in the presence
of random on-site potentials (local magnetic fields) 
featuring long-range spatial correlations 
[see Eq. (\ref{disorder})]
starting from a fully localized spin excitation in the bulk of a polarized background.

\subsection{Ordered chain}

We start off our discussion for the noiseless case, that is $\varepsilon_{i} = \varepsilon$.
The dynamics of entanglement for the uniform chain has already been
studied in a very detailed way in Ref. \cite{amico04}.
For completeness, we now briefly recall the main aspects of it. 
That settles the ground for our following investigation.
%

%Basically, all the input we need to compute the properties 
%discussed in the previous section
%can be extracted from 
Given the unitary evolution, 
$\ket{\psi(t)} =\hat{\mathcal{U}}(t)\ket{\psi(0)}$, 
where $\hat{\mathcal{U}}(t)\equiv e^{-i\hat{H}t}$ is the unitary time-evolution operator.
we have, in terms of the spectral decomposition of the Hamiltonian, 
\begin{equation}
\ket{\psi(t)} =e^{-iHt}\ket{\psi(0)}=\sum_{k}e^{-iE_{k}t}\ket{E_{k}}\bra{E_{k}}\psi(0)\rangle 
\end{equation}
%where $\lbrace E_{k}\rbrace$ and $\lbrace \ket{E_{k}} \rbrace$ 
%denote the eigenvalues and eigenstates of $H$, respectively. 
For a translational-invariant array, the normal modes are well known (plane waves) 
and read 
\begin{eqnarray}
E_{k} &=& 2J\mathrm{cos}(k/2), \\
\label{eigenstate}\ket{E_{k}} &=& \sqrt{\dfrac{2}{N+1}}\sum_{x=1}^{N}\mathrm{sin}\dfrac{kx}{2}\ket{x},
\end{eqnarray}
with $k=2\pi m/(N+1)$ and $m=1,\ldots,N$. 
Given a fully-localized initial state at site $x_{0}$, $\ket{\psi(0)}=\ket{x_{0}}$, 
the evolved state features
coefficients (herein we take $J=1$ or, equivalently, $t\rightarrow tJ $)
\begin{equation} \label{amp}
w_{x}(t) = \bra{x}\psi(t)\rangle = \dfrac{2}{N+1}\sum_{k}e^{i2\mathrm{cos}(k/2)t}\mathrm{sin}(kx)\mathrm{sin}(kx_{0}),  
\end{equation}
where the probability of finding the spin excitation at position $x$ is
simply the absolute square of the above expression, $|\bra{x}\psi(t)\rangle|^2$. 

% FIG 2
\begin{figure}[t!] 
\includegraphics[width=0.42\textwidth]{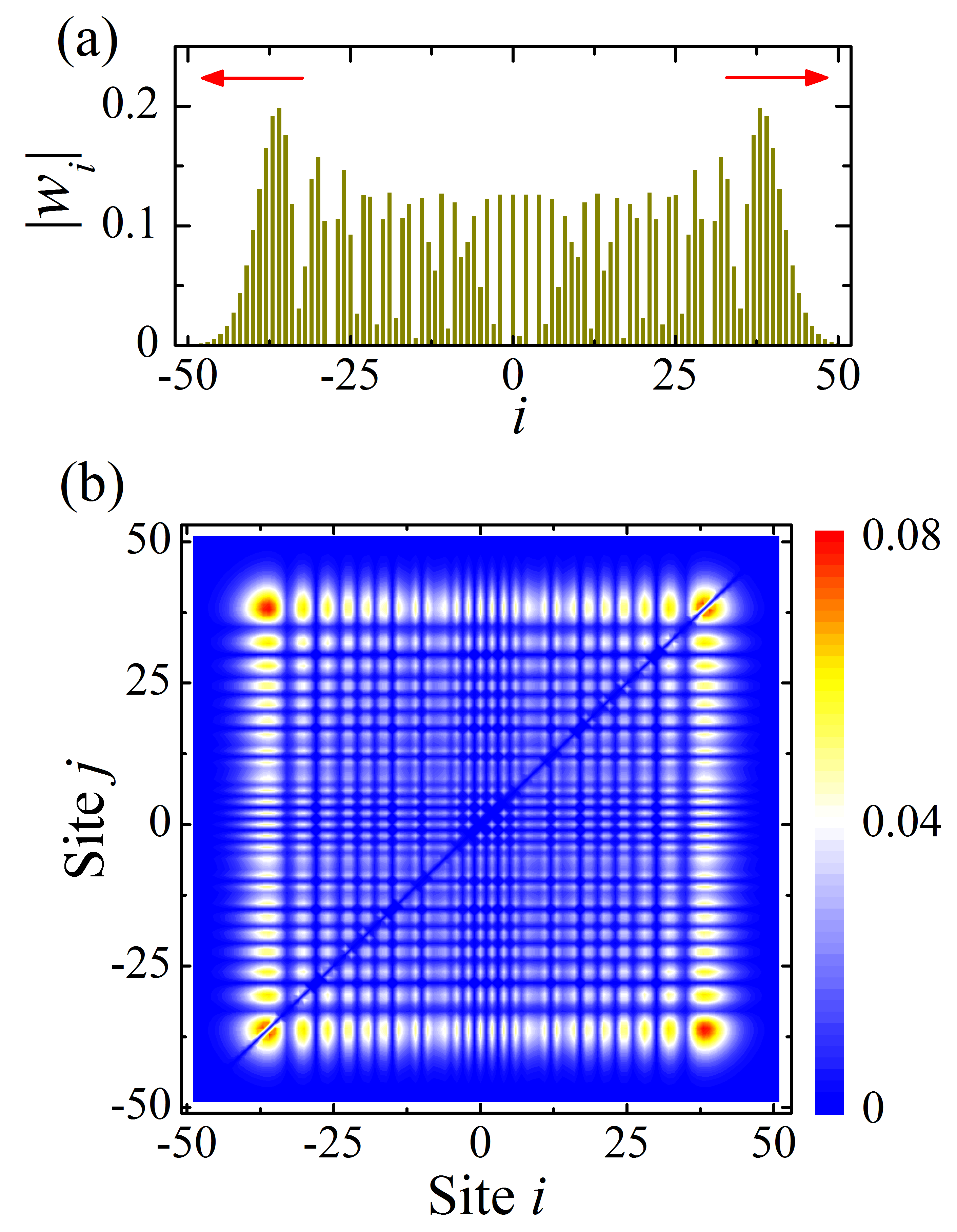}
\caption{\label{fig2} 
(Color online)
(a) Wave-function amplitude distribution for the ordered case when $tJ = 20$ for an excitation initially prepared at site $0$.
It propagates outwards reaching distant sites at $t\approx m$, with $m$ being the distance from the origin. 
(b) Corresponding entanglement distribution profile measured by the concurrence $C_{i,j}$.  Note that the 
``entanglement wave'' mostly involves 
pairwise correlations between
the sites located within the largest wave amplitude and the rest of the chain. Basically, 
this checkerboard pattern is maintained while the evolves and tends to become
homogeneously distributed after a long time. 
Both plots were obtained directly from Eqs. (\ref{c_ij}) and (\ref{bessel}). 
%The concurrence between a given site and itself was set to null. 
}
\end{figure}

In this work, despite considering the spin chain to be essentially finite, we are interested 
in studying the entanglement distribution in the neighborhood of the initial site \textit{before} the excitation reaches the boundaries. Thereby, 
we can effectively work in the thermodynamic limit $N\rightarrow \infty$ where
Eq. (\ref{amp}) takes the convenient form \cite{ben-avraham04, konno05, amico04}
\begin{equation}\label{bessel}
w_{x}(t) = i^{|x-x_{0}|}J_{|x-x_{0}|}(2t),
\end{equation}
%$\tilde{J}_{\nu}(z) = i^{\nu}J_{\nu}(z)$, 
where $J_{\nu}(z)$ is
the $\nu$-th Bessel function of the first kind.
From the properties of the Bessel functions, 
its maximum amplitude decays as $1/\sqrt{m}$ where
$m\equiv|x-x_{0}|$ is the distance from the initial flipped spin. 
This maximum is associated with the \textit{first} root of the derivative of the Bessel function, which we denote by $z_{m}^{*}$. 
%In other terms, $J_{m}(t)$ is maximum 
%when $t=t_{m}^{*}$. 
Using $\mathrm{d}[z^{m}J_{m}(z)]/\mathrm{d}z = z^{m}J_{m-1}(z)$,
one can easily show that 
\begin{equation}\label{timescale}
z_{m}^{*} = m \dfrac{J_{m}(z_{m}^{*})}{J_{m-1}(z_{m}^{*})}
\end{equation}
(note that $\mathrm{d}J_{m}(z)/\mathrm{d}z=0$ at $z=z_{m}^{*}$). 
Straightforward numerical analysis shows that $z_{m}^{*} \approx m$ for the first root.
%This linear relationship embodies a ballistic spreading. 
Therefore, the wave-function reaches a given site 
roughly at $t\approx m/2$ and,
after reaching its first maximum,
it goes on oscillating with the same frequency but 
decaying as $1/\sqrt{t}$.

% Concurrence, symmetry and wave envelope.
Keeping in mind what has been discussed above, 
one may grasp the concurrence behavior right away [see Eqs. (\ref{c_ij}) and (\ref{bessel})].
Basically, the excitation spreads out ballistically from the origin in the form of two
dispersing envelopes as shown in Fig \ref{fig2}(a). 
Entanglement turns out to be concentrated at the front wave packet corresponding to the first local maximum occurring when $t\approx m/2$,
in which the corresponding site becomes \textit{mostly} entangled with its first neighbors as well as with equidistant sites due to the symmetric nature of the wave-function [see Fig \ref{fig2}(b)]. This
very region also gets partially entangled with the remaining sites
wherever there is a non-vanishing wave amplitude. 
The concurrence then goes on decaying with time and
oscillating with a well-defined period \cite{amico04}.
% middle site does not share entanglement; subsequent revivals; oscillating behavior
%
% 

%
% FIG 3
\begin{figure}[t!] 
\includegraphics[width=0.49\textwidth]{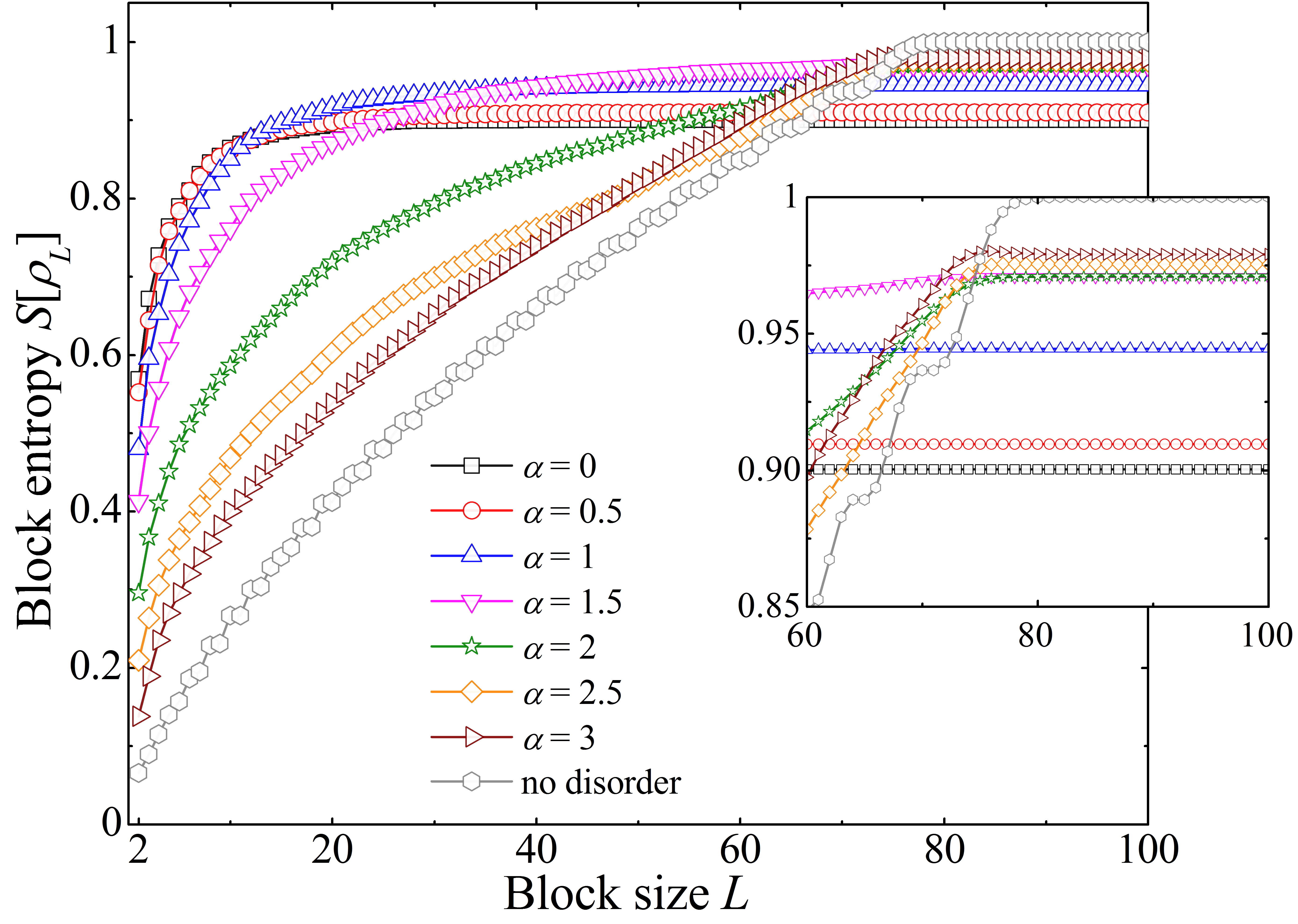}
\caption{\label{fig3} 
(Color online) Entropy for a block of spins, $S[\rho_{L}]$, versus its size $L$
when $tJ = 40$ for many disorder regimes 
provided by $\alpha$
including uncorrelated disorder ($\alpha = 0 $) and 
the noiseless condition. 
Each block involves groups
of spins ranging from ($x_{0}+1$)-th to the $L$-th spin, 
with $x_{0}$ being the initial site (which is out of the block) located in the middle of the chain.
The entanglement saturation threshold sets the extension of the wave-function
in a given time. Note it happens quite rapidly for lower values of $\alpha$.
%point after which
The inset provides a zoomed-in view of the saturation region 
for higher values of $\alpha$.
%highlighting
%how likely the excitation is to be found at $x_{0}$.  
%
Plots were obtained from exact numerical diagonalization of Hamiltonian (\ref{fermionH}) 
for $N = 400$ and $S[\rho_{L}]$ was averaged over $10^{2}$ independent realizations of disorder.
}
\end{figure}

In summary, in a completely uniform chain  the excitation, as well 
as the entanglement, tends to become homogeneously distributed across the chain, for sufficiently long times, due to 
the extended nature of the underlying eigenstates [see Eq. (\ref{eigenstate})]
regardless of the initial conditions. Take, for instance, an initial 
entangled Bell state of the form $\left( \ket{0_{i}1_{j}} \pm \ket{1_{i}0_{j}} \right) \otimes \ket{\mathrm{vac}}/\sqrt{2} = \left( \ket{i}\pm \ket{j}\right)/\sqrt{2}$. 
In this case, the wave-function coefficients [Eq. (\ref{bessel})]
can be expressed as \cite{amico04}
\begin{equation}
w_{\ell}(t) = \dfrac{1}{\sqrt{2}}\left[ J_{\ell - i}(2t) + i^{(i-j)}J_{\ell-j}(2t) \right].
\end{equation}
There are no qualitative differences between 
initializing the system with a single excitation
or a maximally-entangled Bell state of the above form -- the latter case will feature different interference
profiles in between both initial sites depending on the distance between them -- as both 
access the same set of eigenstates during the evolution.
Because of that,  
here we focus on the case of a   
single flipped spin prepared
on a polarized background. 
%(vacuum state in terms of the fermionic description in Eq. xx).
Moreover, this problem has been addressed experimentally \cite{fukuhara13,*fukuhara15} 
using ultracold atoms in optical lattices to study entanglement propagation.

The dynamics discussed above can be seen as a protocol for generating entanglement between distant sites through
the natural evolution of the spin chain \cite{bose03, wojcik05, wojcik07, plenio04, banchi10, *banchi11}, though it gradually becomes weaker
with distance due to dispersive effects (since the chain is uniform).
However, this entanglement may be properly distilled into pure singlets \cite{horodecki98, bose03} thus building up 
resources for quantum teleportation schemes. 
 
\subsection{Disordered chain} 

By adding uncorrelated on-site disorder to the system, the above scenario changes dramatically.
In one-dimensional tight-binding models, the presence of disorder is well-known
for inducing Anderson localization \cite{anderson58} of every eigenstate no matter 
how weak disorder is. 
Each mode 
gets exponentially localized around a site, that is $\langle x\vert E_{k}\rangle \sim e^{-\frac{|x-x_{0}|}{\xi_{k}}}$ 
for a given $x_{0}$, where $\xi_{k}$ 
accounts for the length of localization \cite{thouless74}.
As a consequence, the excitation is unable to
spread out too far from the origin and so entanglement remains
concentrated there for all times [cf. Eq. (\ref{c_ij})].    
On the other hand, by adding correlations in the disorder distribution, particularly long-ranged, 
the emergence of extended states in the middle of the band 
\cite{demoura98,lima02, demoura02, adame03} and their
interplay with localized states beyond the mobility edges bring in very
interesting resources for entanglement distribution as we are going to show now.

% FIG 4
\begin{figure*}[t!] 
\includegraphics[width=0.97\textwidth]{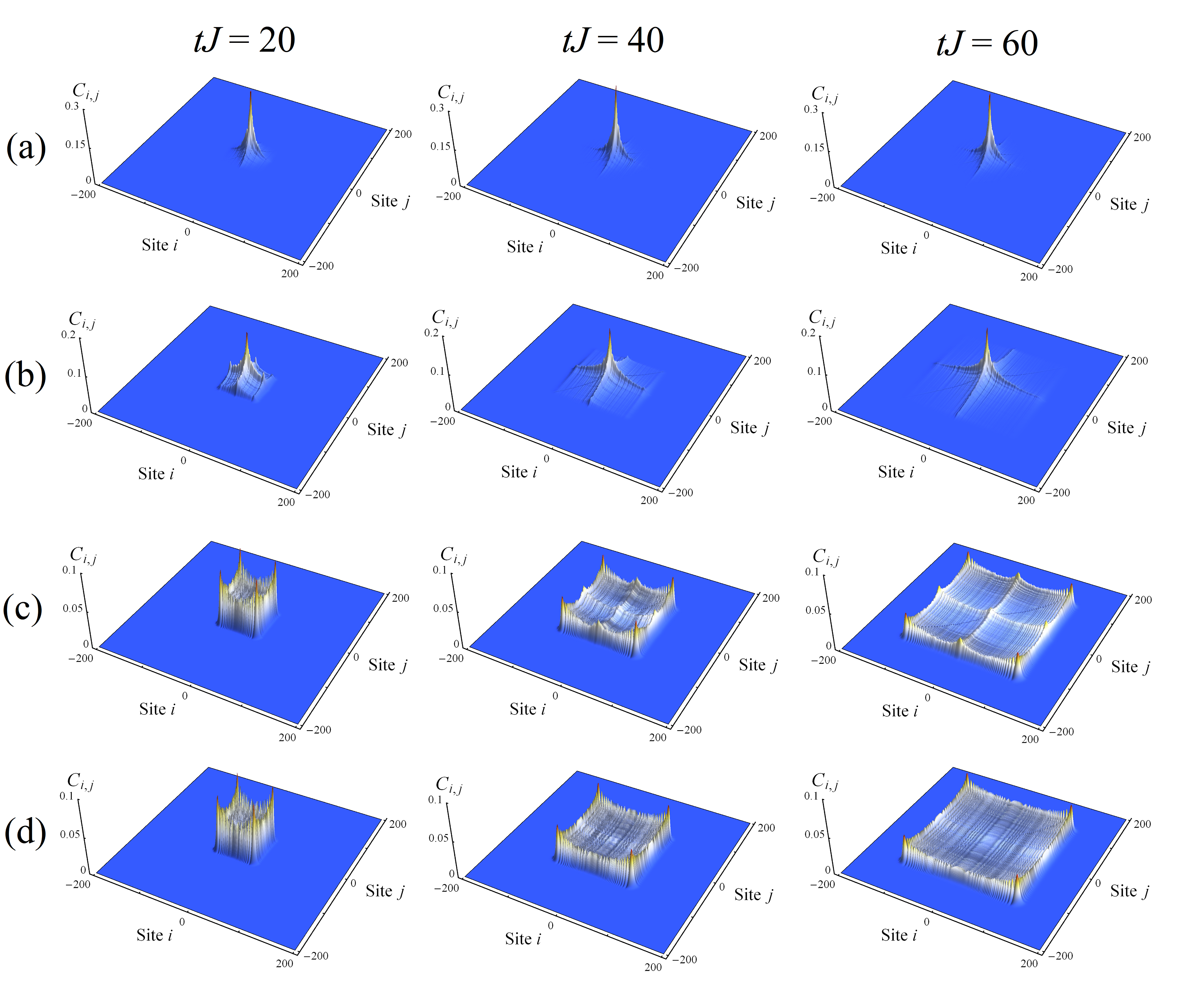}
\caption{\label{fig4} 
(Color online) Snapshots of concurrence distribution 
for (a) $\alpha=0$ (uncorrelated disorder), 
(b) $\alpha=1$, (c) $\alpha=2$, and (d) $\alpha=3$ 
averaged over $10^{2}$ independent realizations of disorder.
The initial state was set in the middle of a chain 
with $N=2048$ sites.
We set 
$C_{i,j} = 0$ for $i=j$.
The first, second, and third columns correspond to times $tJ = 20$, $40$, and $60$, respectively.
Here, we clearly see that long-range correlations in the disorder distribution push
the entanglement wave to reach distant sites, first extending the 
localization length and then setting up genuine 
propagating modes when $\alpha=2$ and higher.
}
\end{figure*}
In order to see how far the initial excitation propagates depending on the degree of correlation $\alpha$ [see Eq. (\ref{disorder})] let us first 
analyze the entanglement entropy for a given block
of spins $\mathcal{A}_{L}$ of size $L$
as defined in Eq. (\ref{entropy}). We initialize the system 
as a single flipped spin located at the middle of the chain (say, the zeroth site $\ket{x_{0} = 0}$)
and evaluate the block entropy for increasing $L$ where each block
begins from the first-neighbor site $x_{0}+1$ and goes forward until the $L$-th site.
The reason for choosing such a partition is the following: 
as the block entropy, Eq. (\ref{entropy}), solely depends on the overall probability
of finding the excitation \textit{inside} the block, 
whenever its
saturation occurs for
a given value of $L$, in a given instant, it means
that the excitation has not yet reached (or never will) the subsequent sites (that is, $x>L$ and $x<-L$). 
Moreover, the saturation value accounts for how much the initial
site is still populated. Note that the wave-function spreads out symmetrically and, as already mentioned, the 
entropy reaches its maximum value, that is 1, when $p = 1/2$.
%

%Discussion of Figure 3
The behavior of the entanglement entropy 
as a function of the block size $L$
is reported in Fig. \ref{fig3} for several disorder configurations (including the noiseless case for comparison).
We stress that every quantity shown in this work is properly averaged 
over about $10^{2}$ \textit{independent} disorder realizations. 
In Fig. \ref{fig3} we 
note that each curve saturates after crossing 
a threshold value for $L$, as mentioned before.
This indicates the spot after which $p = \sum_{i \in \mathcal{A}_{L}}\vert w_{i} \vert^{2}$ is no longer added up. 
%i.e., the excitation has not crossed that size.  
The saturation takes place quite fast, as expected, for the uncorrelated disorder scenario ($\alpha = 0$) in which pure Anderson localization sets in. A similar profile 
is maintained for low degrees of correlation, though for higher $L$.
%it seems that the wave-function extends a bit further. 
%
The behavior suddenly changes for $\alpha = 2$ and above, where the disordered potential landscape is characterized by self-similar persistent increments [see Fig. \ref{fig1}(b)]. 
Now, 
the entropy slowly increases with $L$ indicating that 
extended states are indeed taking part on the evolution. 
Therefore, $\alpha = 2$ marks the transition point between the localized and delocalized regimes \cite{demoura98, adame03}. We will discuss how it affects the distribution of entanglement in a moment. 
Note that the excitation has reached out about the 80-th spin corresponding to the ordered curve, which is the
farthest it can go for $tJ = 40$ [cf. Eq. (\ref{timescale})]. 
Also, 
since in this case the entropy goes to $1$, 
the excitation has almost completely left the initial site.
%there is about nothing left at the initial 
%site 
(see inset of Fig. \ref{fig3}). 
%Note, however, that 
%for higher values of alpha the maximum entropy seems to accumulate    
%so still remaining pretty far from the boundaries.
%

%Discussion of Figure 4
Now, let us take a more detailed view on the way entanglement is distributed between a
pair of spins along the chain in the disordered scenario. Fig. \ref{fig4} shows time snapshots
of the concurrence grid for several values of $\alpha$.   
In the case of uncorrelated disorder [Fig. \ref{fig4}(a)], the concurrence barely propagates, as expected,
since the spin excitation remains strongly localized at the initial site.
The slightest amount of long-range correlations in the disorder distribution allows for
a broader distribution of entanglement. Already for $\alpha = 1$ in Fig. \ref{fig4}(b), 
the concurrence between the initial site and the remaining ones extends further out.
In other words, the localization length effectively increases. Interestingly, a closer look at the first panel of Fig. \ref{fig4}(b) (that is for $tJ = 20$) indicates the 
appearance of a (very) small envelope surrounding the central peak that 
rapidly dissipates with time. This is signaling the emergence of weakly localized modes \cite{demoura98,lima02, demoura02, adame03} although
the dynamics is still ruled by the strongly localized modes. 
%That is why 
%the central spin is able to communicate with
%more distant sites at first place.
%Therefore, in principle, localization is not that bad for entanglement generation
As we saw in Fig. \ref{fig3}, $\alpha = 2$ sets the transition 
to a delocalized-like behavior. In terms of entanglement dynamics [see Fig. \ref{fig4}(c)], the interplay between localized and delocalized states 
keeps the central spin able to correlate with distant sites.  
At the same time
the propagating wave front responsible for that 
also generates entanglement between
%creates its own entanglement network involving
neighboring spins and their equidistant counterpart. 
If we increase the degree of noise correlations, i.e. by increasing $\alpha$, the 
entanglement involving the initial site is substantially decreased [Fig. \ref{fig4}(d)]
and the distribution pattern resembles that of the ordered case shown in Fig. \ref{fig2}(b). 
Indeed, if $\alpha$ is large enough the summation in Eq. (\ref{disorder}) 
reduces to a single cosine function in the asymptotic limit and thus the local magnetic field acquires a smooth periodic behavior in space, thereby suppressing localization. 

% 
% FIG 5
\begin{figure}[t!] 
\includegraphics[width=0.49\textwidth]{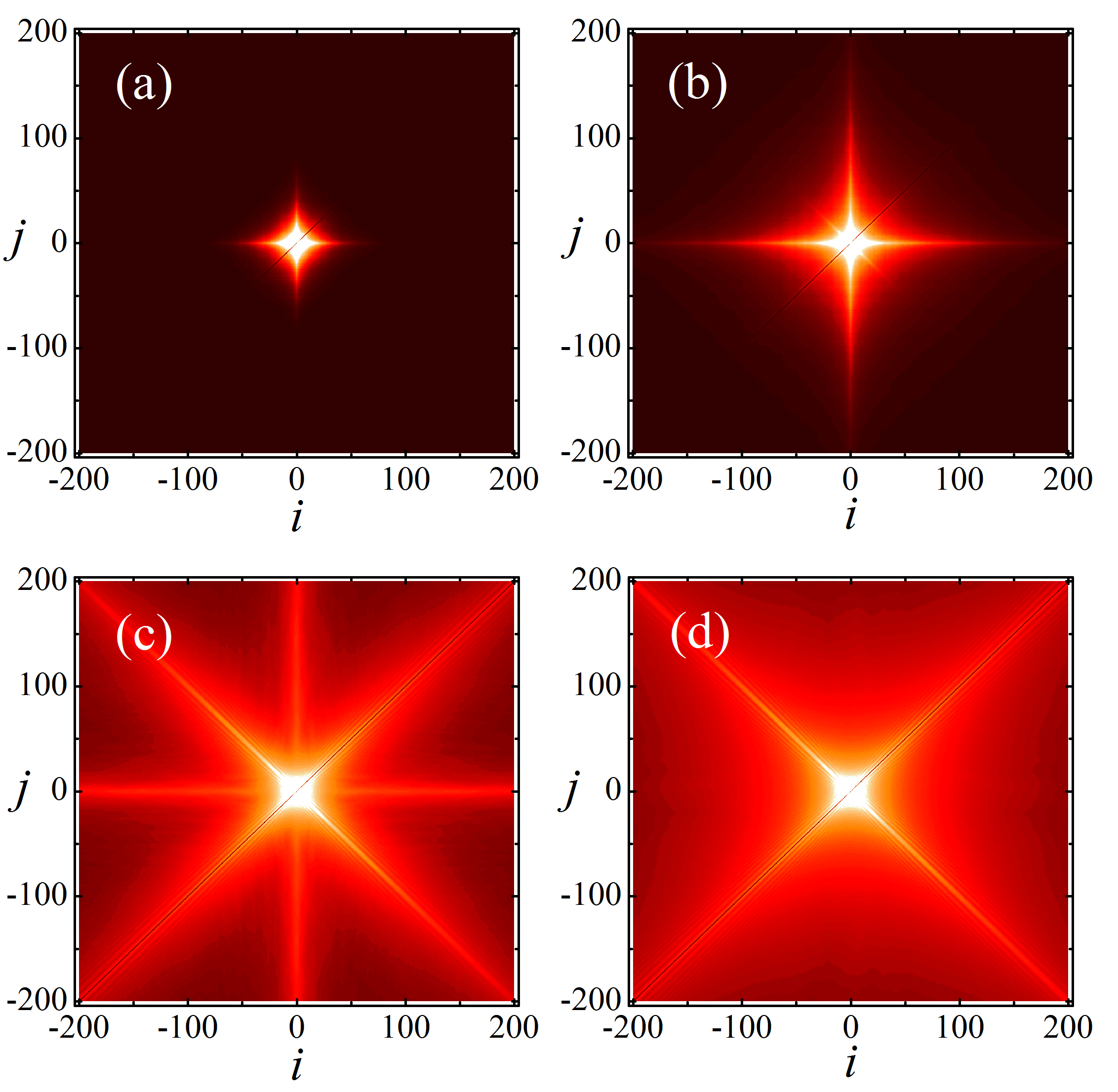}
\caption{\label{fig5} 
Maximum concurrence $C_{i,j}^{\mathrm{max}}$  for (a) $\alpha = 0$ (uncorrelated disorder),
(b) $\alpha = 1.0$, (c) $\alpha = 2.0$, and (d) $\alpha = 3.0$ averaged over $10^{2}$ independent realizations of disorder. The scale goes from $0$ (dark spots) to $0.1$ and above (bright spots).
System's parameters are the same as from Fig. \ref{fig4} and 
the time window considered for $C_{i,j}^{\mathrm{max}}$ was $tJ \in [0,200]$
thus making sure that the entanglement wave has passed through the neighborhood of the initial site (which we denote as the $0$th site for easiness)
without reaching the boundaries of the chain ($N=2048$). 
As we move from uncorrelated disorder towards long-range-correlated disorder, 
there are many configurations available for entanglement distribution. 
%It is worth pointing out
%again that this is coming from very simple situation of a single-flipped spin prepared on
%the fully-polarized state. 
}
\end{figure}
%
%

%Discussion of Figures 5 and 6
In order to show the highest amount of pairwise entanglement
one is able to create  
during the process discussed above,
%of releasing a single spin excitation 
%from the middle of the chain in a 
%disordered magnetic-field landscape, 
in Fig. \ref{fig5} we plot the \textit{maximum} concurrence evaluated
in a given time interval.
That ultimately provides
a bird-eye view on the way entanglement is shared 
among individual spins. In Fig. \ref{fig5}(a) the prominent core witnesses 
the fully-localized nature of the underlying Hamiltonian spectrum.
% which is what
%really happens for $\alpha = 0$. 
For $\alpha = 1$ [Fig. \ref{fig5}(b)], although long-range correlations
are already building up, the latter are not enough for breaking down
the localized behavior, though the concurrence involving the central site
widens out considerably. 
The value $\alpha = 2$ corresponds to the critical correlation degree signalling the emergence of extended states \cite{demoura98}. In this case, entanglement results from the coexistence between order and disorder thus
inducing propagating modes in the dynamics 
while keeping some residual localized-like behavior. 
This results  
in the very interesting pattern shown in Fig. \ref{fig5}(c). 
Extended states then take over the dynamics for higher degrees of correlation as shown
Fig. \ref{fig5}(d) for $\alpha = 3$.

\begin{figure}[t!] 
\includegraphics[width=0.46\textwidth]{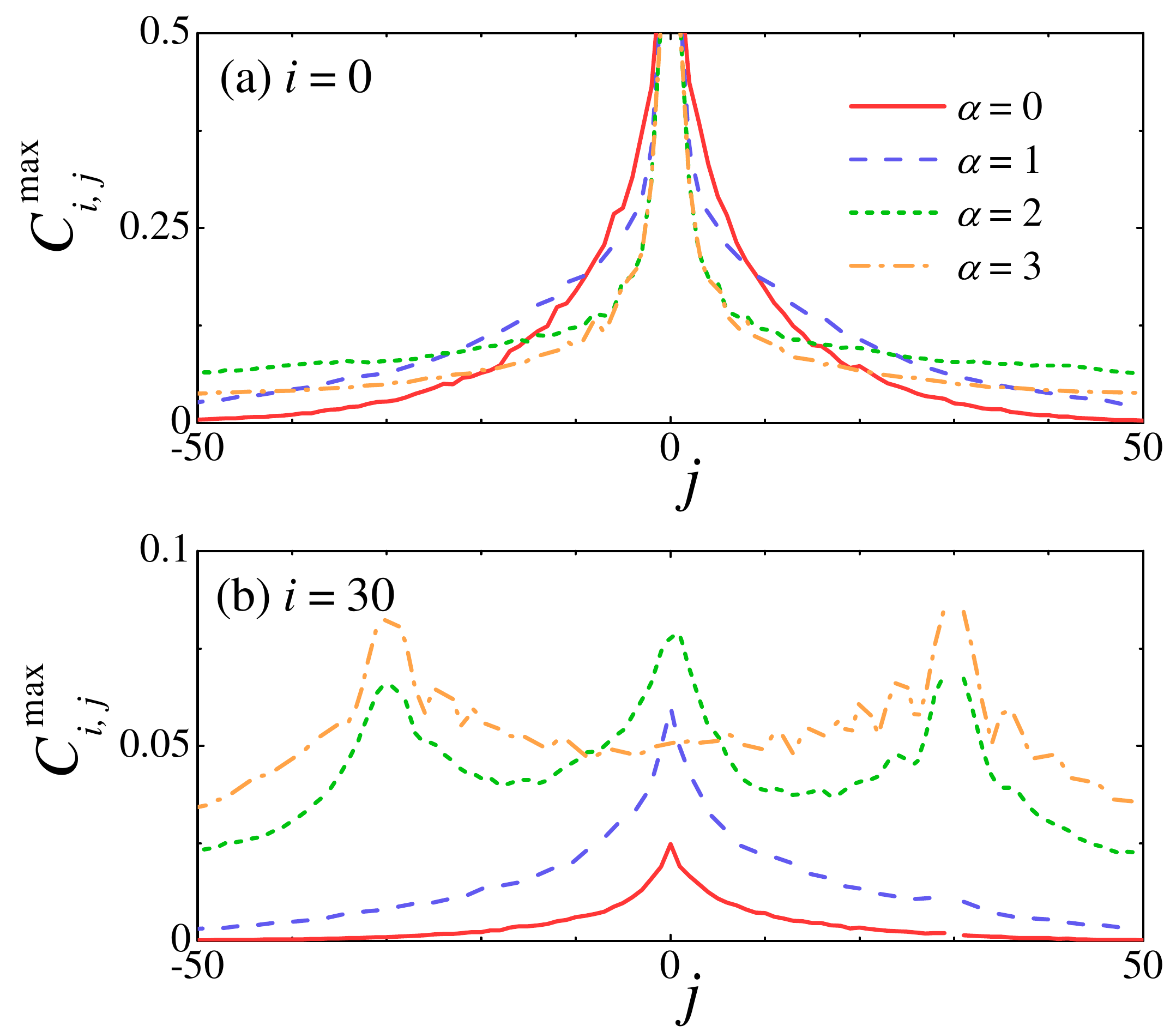}
\caption{\label{fig6} 
Distribution of maximum concurrence $C_{i,j}^{\mathrm{max}}$ for 
fixed (a) $i=0$ (initial site) and (b) $i=30$
and varying $\alpha$. 
Those are basically detailed side views of 
Fig. \ref{fig5} involving a much smaller region. 
Again, the time window was set to
$tJ \in [0,200]$ and $C_{i,j}^{\mathrm{max}}$ was averaged over 
$10^{2}$ independent realizations of disorder. 
%The broken parts along each curve in 
%panel (b) denotes $C_{30,30}^{\mathrm{max}} = 0$. 
}
\end{figure}

Figure \ref{fig6} provides a more detailed side view of the distribution of
$C_{0,j}^{\mathrm{max}}$ [Fig. \ref{fig6}(a)] and $C_{30,j}^{\mathrm{max}}$ [Fig. \ref{fig6}(b)] ranging over
about a hundred sites. Note that the central site is able to establish a higher entanglement with distant sites when $\alpha = 2$, suddenly decreasing when $\alpha=3$.
Also, for $\alpha = 2$ and above, 
note the formation of symmetric peaks accounting for the entanglement 
in between small groups of neighboring spins and their equidistant parts.

% 
% FIG 6

\subsection{Entanglement transmission}

So far we have been discussing the generation and spreading of entanglement
from the origin and analyzing mostly its distribution patterns 
as a function of the degree of correlations in the disorder distribution. 
We now turn our attention to a slightly different problem, 
which is the task of transmitting 
entanglement along the chain \cite{bose03}.
Suppose we prepare a maximally entangled state of the form 
$\ket{\psi(0)}=\frac{1}{2}(\ket{0_{s} 1_{s'}} + \ket{1_{s} 0_{s'}})$
involving spin $s$, belonging to the chain, and an external (uncoupled) one $s'$.
Because of the Hamiltonian dynamics, the component of the wave-function featuring the excitation inside the chain evolves and causes the entanglement, initially present only between site $s$ and $s'$, to get shared among the latter and other spins of the chain.  
Hence, the figure of merit of the entanglement between a given spin $r$ (belonging to the chain) and the external spin $s'$ 
is given by
%between the external and receiver spins,
$C_{r}(t) \equiv \sqrt{2}|w_{r}(t)| = |\langle r \vert \hat{\mathcal{U}}(t) \vert s \rangle|$.
This can be worked out by using the same dynamics 
discussed previously.
%During the propagation, the mobile spin is also responsible for 
%sharing entanglement across the chain. 

It should be noted that
single-particle states saturates the Coffman-Kundu-Wootters conjecture \cite{coffman00} (see also \cite{amico04}), which means that 
all the entanglement present in the system 
is encoded by pairwise correlations only (there is no higher-order entanglement).
As a consequence, it is easy to check that $\sum_{r}C_{r}^{2} = 1$. 
Hence, we can use this fact to 
build up proper entanglement transmission and reflection coefficients of the form \cite{apollaro06}
\begin{equation}\label{TR}
T = \lim_{t \rightarrow \infty} \sum_{r>r_{0}}C_{r}^{2}(t), \,\,\,\,\,\,\, R = \lim_{t \rightarrow \infty} \sum_{r<r_{0}}C_{r}^{2}(t),
\end{equation}
where $r_{0}$ denotes a particular reference spin for which the above coefficients
stand for. 
%We stress that, just like in previous sections, 
%we will perform the calculations for each sample separately and then average them out by %the number of realizations for a given disorder configuration. 

\begin{figure}[t!] 
\includegraphics[width=0.46\textwidth]{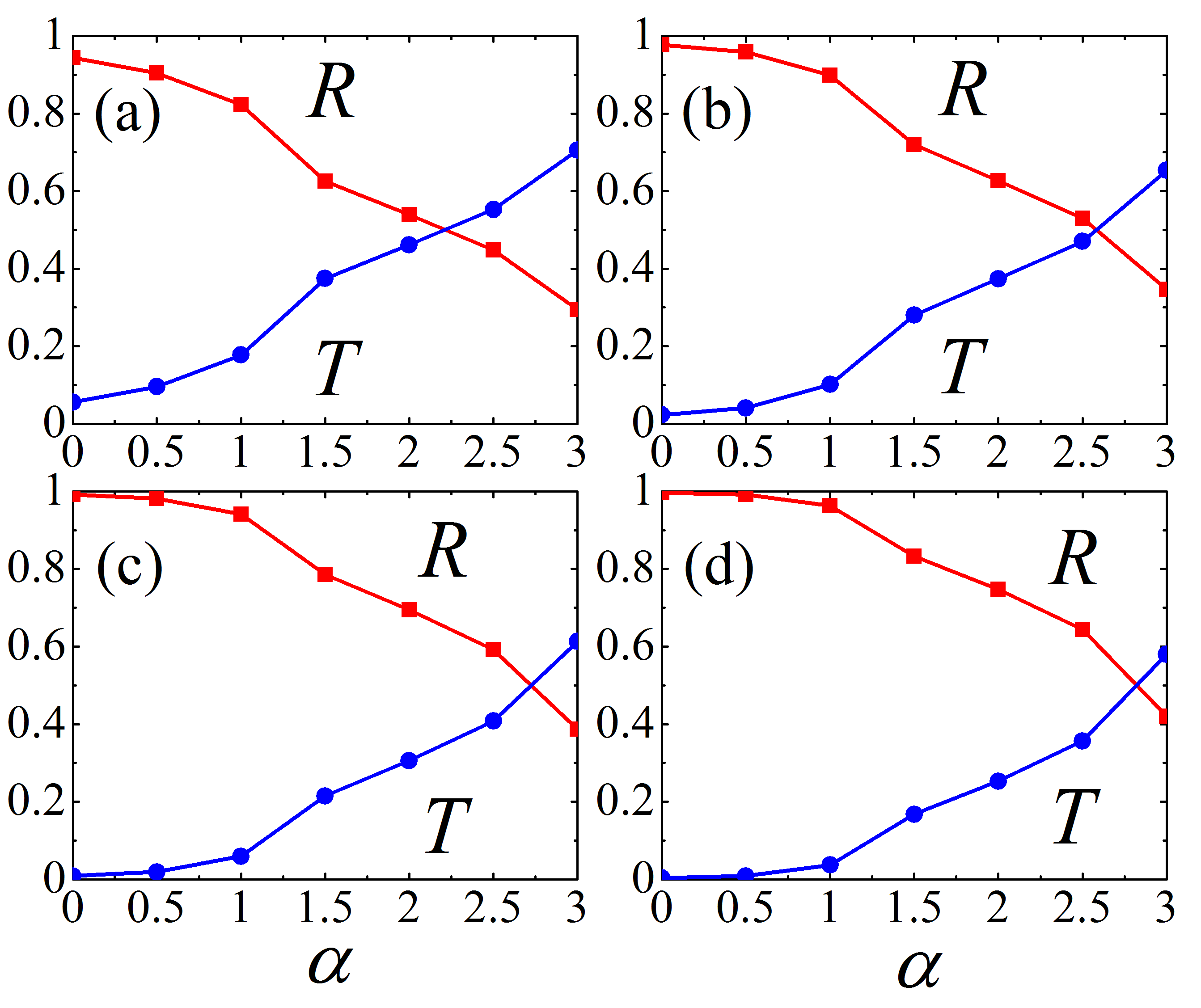}
\caption{\label{fig7} 
Entanglement transmission and reflection coefficients [Eq. (\ref{TR})] versus $\alpha$
for 
(a) $r_{0} = 20$, (b) $r_{0} = 30$, (c) $r_{0} = 40$, and (d) $r_{0} = 50$.
The initial state was a maximally-entangled pair
$\ket{\psi(0)}=\frac{1}{2}(\ket{0_{s} 1_{s'}} + \ket{1_{s} 0_{s'}})$ 
with the sender prepared in the first site ($s=1$) in a chain of size $N=400$ (note that now, for convenience, we are numbering each spin regularly from 1 to $N$).
The calculations were performed 
without letting the wave-function reach the other end of the chain
and the coefficients were 
evaluated at times when $C_{r}^{2}(t)$ (averaged over $10^{2}$ samples) achieved a stationary behavior in order to account for the long-time regime.
}
\end{figure}

Figure \ref{fig7} shows the transmission and reflection coefficients as a function of the degree of correlations $\alpha$ for several reference spins $r_{0}$. 
The initial state was 
$\ket{\psi(0)} = \frac{1}{2}(\ket{s}+\ket{s'})$, 
with $s$ denoting the first site in a chain of $400$ spins. We stress that all the calculations were performed before the wave-function reached the other boundary and $T$ and $R$ were 
evaluated at times when $C_{r}^{2}(t)$ (averaged over the number of realizations) achieved a stationary behavior.
Unlike the $\alpha = 0$ (uncorrelated disorder) case, where
the transmission coefficient is, for all practical purposes, negligible already 
when considering the propagation of entanglement across 
the $20$th site of the chain [Fig \ref{fig7}(a)],  
there is a quite significant transmission gain
%clearly a major difference in the transmission scenario
when $\alpha \neq 0$. In Fig \ref{fig7}(a), for instance, 
it goes from $T \approx 0.05$ to $T \approx 0.7$ when $\alpha=3$. 
Even though 
$T$ slowly diminishes [see Fig \ref{fig7}] 
as we get more distant from $s$, i.e., by increasing $r_{0}$ in Eq. (\ref{TR}),
we nevertheless see a monotonic increase of the transmission coefficient by increasing the disorder correlations.
Furthermore, we note 
that $T$ surpasses $R$ when 
$\alpha>2$ thus once again revealing the presence
of delocalized states in the spectrum \cite{demoura98, adame03}.

% FIG 7

\section{\label{sec5}Concluding remarks}

In our work, we addressed the problem of creating and distributing entanglement
in disordered 1D spin chains
by means of the time evolution of a single flipped spin prepared on
the fully-polarized state. 
We considered on-site diagonal disorder (that is, on the local magnetic field distribution) 
with long-range spatial correlations
resulting from the power-law nature of the spectral density, $S(k) \propto 1/k^{\alpha}$.
A rich variety of dynamical regimes
for the entanglement we reported here has been generated by
varying a single parameter, namely the degree of correlations 
in the on-site magnetic field disorder distribution,
$\alpha$.
This parameter is known for dictating the appearance of extended states 
in the middle of the band \cite{demoura98, lima02, demoura02} thus allowing for 
sending single-particle pulses outwards
while still maintaining part of the wave-function amplitude at the center of the chain.
We also studied the propagation of entanglement 
from an initial maximally-entangled Bell state through the chain and found that there is a significant improvement in the transmission coefficient by increasing $\alpha$. 

%Creating extended states in noisy channels can also be 
%very useful for some QST protocols \cite{yao11}.
%Therefore, disorder might not always be harmful 
%as one usually thinks.
% 
%...where the random local magnetic fields
%feature long-range internal correlations with power-law 
%spectral density. 

In general, correlated disorder 
may naturally be present 
in solid-state devices due to the 
lack of full experimental control over the system 
itself as well as over the surroundings. 
However, as long as internal correlations of a specific kind
generate the desired eigenfunctions profile, it should be much more 
convenient to cope with disorder than fighting against it, since 
the latter strategy may demand more resources.  
%a high degree of precision and control over the environment.
In other words, when designing a given quantum information processing protocol to be realised in a disordered system, 
one could think of increasing the amount of correlations of such a disorder, instead of trying to get rid of it.
%setting the parameters following predefined rules instead of 
%predefined specific values. 
%That opens up a prominent path in the design of quantum computing architectures.

Although we have considered only diagonal disorder, a similar behavior is expected
for off-diagonal fluctuations, i.e., a disordered set of exchange interactions albeit
some quantitative differences \cite{lima02, demoura02}. 
%
%Moreover, the external magnetic field
%stands out as a knob for local manipulation
%once the chain is already operating [\textbf{is it, TONY??}].
% 
In principle, our findings can be probed in, e.g., ultra-cold atoms in optical lattices 
\cite{fukuhara13, *fukuhara15} in which great advances 
such as single-site addressing \cite{weitenberg11} has been achieved .

Further extensions of our work include investigating the role of 
internal correlations in 
disordered channels for high-fidelity state transfer based on
weakly-coupled communicating parties \cite{wojcik05, yao11, almeida16}. In these models, even though the bulk of the chain is 
usually weakly populated during the transmission process, the presence
of extended eigenstates in the bulk is crucial to support long-distance 
communication protocols.
%\cite{wojcik05,wojcik07, huo08, kuznetsova08, lorenzo13, almeida16}.

\section{acknowledgments}

This work was partially supported by CNPq (Conselho Nacional de Desenvolvimento Cient\'{i}fico e Tecnol\'{o}gico), Grant No. 152722/2016-5,
CAPES (Coordenação de Aperfei\c{c}oamento de Pessoal de Ensino Superior,  FINEP (Financiadora de Estudos e Projetos), and FAPEAL (Funda\c{c}\~{a}o de Amparo \`{a} Pesquisa do Estado de Alagoas). 
 T.J.G.A. acknowledges support from the Collaborative
Project QuProCS (Grant Agreement 641277).

%Here we have shown a special case, i.e., other kinds of correlations short- or
%long-ranged might also be explored. 
%
%Our point is: it is difficult to fight against 
%disorder and setting specific parameters with extremely high precision
%through the whole chain. 

%I took this from demoura98: 
%These have the feature of displaying well defined power-
%law  spectral  densities  in  contrast  to  real  
%correlated  sequences which exhibit a 
%noisy power-law spectrum. We
%expect that the present filtering of the noise in the amplitudes 
%of the Fourier components of the potential does not
%include or remove any relevant feature associated with the
%underlying correlations.

%Our results are applicable in any physical model isomorphic to the XX spin-1/2
%Hamiltonian (e.g., coupled-cavity systems \cite{almeida13})
%in the right configuration.  

%OUR RESULTS HERE SET ANOTHER WAY OF THINKING ABOUT CARRYING OUT
%QIP PROTOCOLS IN IMPERFECT SETTINGS.   

%\bibliography{refs}


\end{document}